\documentclass[aip,reprint]{revtex4-1}
\usepackage{graphicx}
\usepackage{color}
\usepackage[linkcolor=blue]{hyperref}
\begin{document}

\title{Theoretical Reconstruction of Realistic Dynamics of Highly Coarse-Grained \textit{cis}-1,4-Polybutadiene Melts}

\author{I. Y. Lyubimov, M. G. Guenza\footnote{Author to whom correspondence should be addressed. Electronic mail: mguenza@uoregon.edu}}
\affiliation{Department of Chemistry and Institute of Theoretical Science, University of Oregon, Eugene, Oregon 97403}
\date{\today}

\begin{abstract}
The theory to reconstruct the atomistic-level chain diffusion from the accelerated dynamics that is measured in mesoscale simulations of the coarse-grained system, is applied here to the dynamics of \textit{cis}-1,4-Polybutadiene melts where each chain is described as a soft interacting colloidal particle. The rescaling formalism accounts for the corrections in the dynamics due to the change in entropy and the change in friction that are a consequence of the coarse-graining procedure. By including these two corrections the dynamics is rescaled to reproduce the realistic dynamics of the system described at the atomistic level. 
The rescaled diffusion coefficient obtained from mesoscale simulations of coarse-grained \textit{cis}-1,4-Polybutadiene melts shows good agreement with data from united atom simulations performed by Tsolou et al. The derived monomer friction coefficient is used as an input to the theory for cooperative dynamics that describes the internal dynamics of  a polymer moving in a transient regions of slow cooperative motion in a liquid of macromolecules. Theoretically predicted time correlation functions show good agreement with simulations in the whole range of length and time scales in which data are available. 
The theory provides, from data of mesoscale simulations of soft spheres, the correct atomistic-level dynamics, having as solo input static quantities.
\end{abstract}

\maketitle

\section{Introduction}
Computer simulations of macromolecular liquids described at the atomistic level are extremely useful because they bridge information between the microscopic molecular structure of the polymeric system, at the given thermodynamic conditions, and its macroscopic properties, such as viscosity, diffusion, and dynamical-mechanical response, which are observed experimentally.\cite{Binder,Frenkel}
It is unfortunate that, despite the progress already occurred in the computational hardware and software, the study of the dynamics of polymeric liquids by atomistic computer simulations is still limited by the impossibility of simulating polymer dynamics in the wide range of time and length scales relevant for these systems. It is known that long simulation trajectories deteriorate with time, following a law defined by the Lyapunov exponent characteristic of the system, and the resulting long-time dynamics is affected by errors.\cite{shadowing,Tisley}

Atomistic simulations of polymeric liquids are limited either in the length and number of macromolecules, or in the maximum timescale that can be reached. The long timescale is a regime of great interest for these systems, because of the relevance of dynamical-mechanical properties of macromolecules in that regime for their industrial applications. Because the limitation concerns only the region of long-time, large-scale properties, it is possible to shift the focus of a simulation to this region of interest by the reduction of the simulated degrees of freedom through the averaging of local scale properties, i.e. adopting a so-called ``coarse-graining" procedure. 

While simulations of coarse-grained systems afford to represent larger length- and longer time-scales than atomistic simulations, they also are marred by the shortcoming that the measured dynamics is unphysically fast. This is the advantage that allows one to perform simulations in the long-time regime, but it is also a problem because the dynamics in the simulation becomes too fast and measured diffusion coefficients are too large. Depending on the level of coarse-graining the dynamics can become orders of magnitude faster than the atomistic dynamics.\cite{ShortPaper,LongPaper} 

Being aware that the dynamics has to be rescaled to recover the correct timescale,\cite{Kroger,Kroger1,Voth} one resorts to the most conventional method, which is to build a calibration curve calculated  by superimposing the long-time dynamics of the coarse-grained and atomistic simulations.\cite{Padding,Klein} The hope is for this calibration curve to be transferable to other systems, or to be identical, and so applicable, for thermodynamic conditions close to, but different than the phase point for which the curve was optimized in the first place. Otherwise, if an atomistic simulation has to be ran every time we need to rescale a coarse-grained simulation there is no advantage in running simulations of coarse-grained (CG) systems. The rescaling factor of the conventional calibration curve is, however,   purely a numerical correction. There is no physical motivation for one to assume that this correction has to be identical for any system or thermodynamic condition different from the ones in which it was optimized. Given that the effective potentials between coarse-grained units are free energies, and as such they are parameter dependent, it is very unlikely that numerically optimized corrections will be identical for different systems. However, if the calibration curve is applied for conditions of temperature, density, and degree of polymerization, very close in the phase diagram to the ones in which the calibration curve was originally optimized, there is hope that the error is very contained. Alternatively, in the case that very limited levels of coarse-graining are applied, e.g. combination of a few atoms to obtain a new coarse-grained unit, the error incurred in using the calibration curve in different conditions, can be small enough to be within the numerical error of the calculation. This is in fact the reason for the observed good agreement in the rescaled dynamics of systems with contained level of coarse-graining.\cite{polymer,pb,Vagelis} 

The goal of this paper is orthogonal to the more conventional calculation of an optimized calibration curve for systems with limited level of coarse-graining. In fact our goal is to investigate the physical motivation that leads to the accelerated dynamics in first place and to provide, through our analysis, a theoretical approach to calculate, from first principle equations, the correction factor that needs to be applied to the fast dynamics measured in coarse-grained simulations to recover the slower and more realistic dynamics of the atomistic description. We study the most extreme level of molecular coarse-graining, where one polymer chain is represented as a soft sphere, because i) the description is analytical allowing the formal study of the rescaling problem, ii) it provides a solid test of the reconstruction procedure because possible errors would be maximized, iii) this level of coarse-graining is associated with the largest computational gain, so it is important.  

Recently we have proposed an original procedure to rescale the dynamics measured in simulations of coarse-grained polymer melts.\cite{ShortPaper, LongPaper} The procedure we proposed does not require to run atomistic simulations to calibrate the dynamics of the coarse-grained simulation. Once one parameter, specifically the diameter of the hard-core monomer potential, is fixed, the rescaling factors are fully determined. The advantage of this method is clear, given that the coarse-grained simulation can reach larger-scale longer-time properties than the atomistic simulation. Because no atomistic simulations are needed, the measured dynamics can be directly rescaled to obtain the dynamics of the real system we wanted to study.

The method has been formally derived, and then applied to systems of polyethylene (PE) melts. The choice of PE as a test system was motivated by the wealth of experimental and simulation data available in the literature. We have shown that the proposed theory is able to predict diffusion coefficients in agreement with both atomistic simulations and experiments for systems in both the unentangled and  entangled regimes, for a range of temperatures and densities. 
For entangled system the approach of dynamical reconstruction includes an extra loop in the calculation of the friction coefficient, which accounts for the fact that in real systems described at the atomistic level, the dynamics is slowed down by the presence of entanglements. We hypothesized for the entanglements to relax in time following the same dynamical mechanism of single chain interdiffusion. In this paper entangled chain dynamics is not investigated as most of the samples in the atomistic simulations are either unentangled or in the region of crossover from the unentangled to the entangled regimes.

One of the questions that still needed to be answered was if our method was able to generate high quality predictions also for macromolecular liquids with a more complex monomeric structure than the simple polyethylene. A first study of this issue is presented in this manuscript which investigates the agreement between theoretically predicted diffusion coefficients for \textit{cis}-1,4-Polybutadiene (PB) chains and data from molecular dynamics simulations performed by Tsolou, Mavrantzas and coworkers.\cite{Tsolou,Mavrantzas1,Mavrantzas2} The approach is completely general and can be applied to polybutadiene chains with different tacticity. For these systems the difference in local chain semiflexibility, which corresponds to different overall chain dimension, i.e. a different radius-of-gyration ($R_g$) and statistical segment length ($l$), ultimately leads to a different diffusion coefficients for polybutadienes with different tacticity. In this paper we limit our calculations to \textit{cis}-1,4-Polybutadiene samples because those are the samples investigated by atomistic simulations and provide a effective test of our procedure for dynamical reconstruction.

Our rescaling method considers two contributions emerging from the consequence of applying the coarse-graining process and derives the theoretical corrections that need to be applied to the dynamics of the coarse-grained description to recover the modalities of the original atomistic system. 
To calculate the analytical correction we adopt  for the atomistic descriptions a simple bead-and-spring description of the chain where each PB polymer of $N$ monomeric units is represented as a collection of $N$ beads connected by semi flexible springs of length $l=\sqrt{6/N} R_g$, and the overall chain dimension is defined by the radius-of-gyration of the molecule, $R_g$. This model is a Rouse approach modified to include chain semiflexibility and has been shown to represent well the diffusion of polymer melts in the unentangled regime.\cite{DoiEdw} From the comparison of this model and the soft sphere representation of the chain, it is possible to evaluate the two contributions that are responsible for the accelerated dynamics, and which have to be included to correct the measured dynamics and bring it to the realistic values of the atomistic (here bead-and-spring) description.  

The first correction emerges from the consideration that due to the process of averaging the intramolecular degrees of freedom, the coarse-grained system experiences a change in its entropy, because a number of local states are neglected. While the atomistic system devotes time to explore local configurations, the coarse-grained system doesn't. To recover the correct dynamics those states need to be included \textit{a posteriori} in the form of an entropic contribution to the free energy that rescales the time measured in the CG simulations. 

The second correction comes from the change of the ``shape" of the molecule once it is coarse-grained. Macromolecules described at the atomistic level become represented by chains of effective atoms, or as bead-and-springs, chains of soft blobs, or even each molecule as a soft sphere. In all these multiple forms the surface of each molecule available to the surrounding molecules, i.e. the ``solvent", is different. The hydrodynamic radius, $r_{hydr}$, of the coarse-grained unit in each description is different and so is its friction coefficient, $\zeta$, defined by Stokes' law as $\zeta=6 \pi \eta_{s} r_{hydr}$, where $\eta_{s}$ is the viscosity of the medium. By solving the memory function for the friction coefficient in each description we are able to calculate the correction factor that has to be applied to each friction coefficient to recover the description at a different level of coarse-graining.  

In this paper our rescaling method is first briefly summarized  and then applied to study the dynamics of cis-1,4-PB and compared with simulation data by Tsolou et.al.\cite{Tsolou} who also investigated the effect of temperature and pressure on the simulations.\cite{Mavrantzas1,Mavrantzas2} 
The global dynamics of the polymer is represented by the diffusion coefficient and the decay of the self-correlation for the end-to-end molecular vector. These dynamical quantities are reconstructed using the method by Lyubimov at al.,\cite{ShortPaper,LongPaper} from the information contained in the trajectories of the mesocale simulations of the coarse-grained polymer melts and show good agreement with the data from united atom simulations. For the internal dynamics the theory for cooperative motion\cite{Marina,MGuenza1,Richter} is used to predict monomer dynamics and the dynamical structure factors, which compare well with the simulations. If atomistic simulations are not available, the theory of dynamical reconstruction is able to predict the correct dynamics from the rescaling of the mesoscale simulations having as an input the polymer radius-of-gyration, which can be calculated from simple structural models for polymers. The accuracy of adopting a freely rotating chain model for the calculation of the polymer radius of gyration is discussed in the last section. A brief summary concludes the paper.

\section{Rescaling of the free-energy and rescaling of the friction}
In this section we briefly present an overview of the main steps in our procedure. A complete and detailed presentation of the same has been already published\cite{ShortPaper,LongPaper} and will not be repeated here.  We start from the consideration that the procedure of coarse-graining corresponds to eliminating some internal degrees of freedom, by combining groups of atoms into one effective CG unit. Specifically, our rescaling theory has been developed for an extended level of coarse-graining where intramolecular coordinates are fully averaged out and the polymer is represented as an isotropic sphere centered on the center-of-mass (com) of the macromolecule. There are some advantages in the choice of this description. First, the CG representation is fully isotropic, and even if it is known that the shape of a polymer is closer to an ellipsoid than to a sphere,\cite{DoiEdw} the total correlation function of the polymeric liquid, i.e. the structure of the polymeric liquid, statistically averaged over all the possible configurations is well reproduced by this model; second the formalism is analytical, which allows us to derive formally many of the physical quantities of interest for both the static and the dynamic properties of the coarse-grained system as a function of the atomistic description; third all the consequences of the coarse-graining procedure are enhanced with this extreme level of coarse-graining and are easier to study in such a description. If the procedure adopted were not precise, the error would be maximized because of the high level of coarse-gaining; the fact that the theory, which is predictive, is found to be in quantitative agreement with both simulated and experimental data is encouraging.\cite{ShortPaper,LongPaper} 

One disadvantage of having such a coarse-grained representation of the molecule is that no information is collected from the mesoscale simulation on the internal dynamics of the polymer. In this way, the soft sphere representation only provides information on the diffusion coefficient of the molecule. However, we show in this paper that from the knowledge of the diffusion coefficient and the related monomer friction coefficient for unentangled chains it is possible to recover correctly the internal dynamics of the chain by applying our theory for the cooperative dynamics of a group of interacting chains, which calculates the single chain dynamics for a macromolecule whose dynamics is coupled by the presence of intermolecular forces with the other interpenetrating macromolecules in the liquid.

In every CG model, the averaging of the intramolecular degrees of freedom leads to a speeding up of the dynamics. When a polymer is represented as a soft sphere the dynamics in the mesoscopic simulation is orders of magnitude faster than in the atomistic description. One of the reasons for this acceleration is the fact that while a macromolecule needs to explore a large number of internal chain configurations for each position of the center-of-mass, the soft sphere instead is free to move in the three-dimensional space undergoing Brownian motion.\cite{entropy1,entropy2} To take into account the missing contribution of the time necessary to explore the internal degrees of freedom, the time measured in the mesoscale simulation of the soft sphere is multiplied by a rescaling factor that is calculated as the ratio of the energy due to the internal degrees of freedom in the two representations, i.e an atomistic and a soft sphere representations. For the atomistic representation we adopt an analytical bead-and-spring model, as previously described.\cite{ShortPaper,LongPaper} This model has been shown to provide a quantitative description of the dynamics for polymeric liquids and for proteins in theta solvent.\cite{Richter,BJ}

The first correction term is calculated from the definition of the internal energy for a simple bead-and-spring model and for the soft sphere CG model. The mesoscale (MS) molecular dynamic (MD) simulations are performed using a potential expressed in dimensionless units, which is a common procedure. The time from the simulations has to be properly normalized  so that the dimensionless simulation time from MS simulations, $\tilde{t}$, once dimensionalized and rescaled reads as

\begin{equation}
t=\tilde{t} R_g \sqrt{\frac{m}{k_BT} \frac{3}{2} N} \ ,
\label{EQ:time}
\end{equation}

\noindent where $m$ is the molecular mass of one chain and $R_g$  is the radius-of-gyration of the molecule, which is an input quantity in our approach. The internal degree-of-freedom averaged out in the coarse-grained description are accounted for by introducing the $\frac{3}{2}N$ factor where $N$ is the number of beads. 

The second rescaling is due to the change in the shape of the molecule represented in the two levels of coarse-graining, and so the effective friction in the two representations.
In the long time the com mean-square displacement is related to the molecular diffusion coefficient as 

\begin{equation}
\langle \Delta R^2(t) \rangle = 6\,D\,t \ ,
\end{equation}

\noindent where $D=k_B T/(N\zeta_m)$ is the diffusion coefficient of the system, which has to be calculated. If we use as a starting point the diffusion coefficient measured in the mesoscale simulation $D^{MS}=k_B T/\zeta_{soft}$, the mean square displacement in the rescaled formalism is give by

\begin{equation}
\langle \Delta R^2(t) \rangle = 6\,D^{MS} \frac{\zeta_{\text{soft}}}{N \zeta_m}\,t \ ,
\label{EQ:rescaling}
\end{equation}

\noindent where the problem reduces to the calculation of the friction coefficient in the monomer and soft-sphere representations. The friction coefficients are calculated from the solution of the memory function in the two representations (bead-and-spring and soft-sphere).

The formalism presented so far, adopts a Rouse-like description of the single chain diffusion for the atomistic level representation of the system. At the coarse-grained level the chain is represented as a soft-sphere. Eq.\ref{EQ:rescaling} shows how the diffusion coefficient measured in a mesoscale simulation of a liquid of soft-spheres, $D^{MS}$, needs to be rescaled to give the mean-square displacement of the real chain, $\langle \Delta R^2(t) \rangle$, as a function of the rescaled time, $t$. The contributions that are still unknown are the friction coefficients in the atomistic and coarse-grained descriptions, which can be solved analytically starting from the definition of the memory function.\cite{Zwanzig}

For a bead-and-spring model the monomer (bead) friction is given by:

\begin{eqnarray}
\zeta_{m} & \cong & \frac{1}{N} \sum_{a,b=1}^{N} \int_0^{\infty} d \tau \Gamma_{a,b} (\tau) \ .
\label{EQ:zetamo}
\end{eqnarray}

\noindent where the memory function is defined as\cite{Schweizer}

\begin{eqnarray}
\Gamma_{a,b}(t) &\cong & \frac{\beta}{3} \rho \int d\mathbf{r} 
\int d\mathbf{r'} g(r) g(r') F(r) F(r')\, \mathbf{\hat r} \cdot \mathbf{\hat r'}\nonumber \\
&&\times\int d \mathbf{R} \, 
S^Q_{a,b}(R;t) S^Q(|\mathbf{r}-\mathbf{r'}+\mathbf{R}|;t) \, ,
\label{EQ:MFMON}
\end{eqnarray}

\noindent where $\beta = 1/k_B T$, $\rho$ is the monomer density and $\mathbf{\hat r}$, $\mathbf{\hat r'}$ are the unit vectors characterizing the direction of the forces acting on monomers $a$ and $b$, respectively. $S^Q(r,t)$ is the projected dynamic structure factor, which includes both intra- and inter-molecular contributions, i.e. incoherent and coherent scattering, and is approximated by its non-projected form as $S^Q(r,t) \approx S(r,t)$, which is an acceptable approximation when the Langevin equation is expressed as a function of the slow variables.\cite{LongPaper}

In Eq.(\ref{EQ:MFMON}) $g(r)$ is the monomer pair distribution function of the molecular liquid, and $F(r)$ is the force due to the effective potential between two monomers, obtained from the solution of the Ornstein-Zernike (OZ) equation by applying the Percus Yevick closure. The monomer potential, which in the atomistic-level simulation is a Lennard-Jones potential, is approximated here by an effective hard-sphere with an effective diameter, $d$, to mimic the properties of the L-J potential.  All the physical quantities that appear in the equation are known and the analytical expression for the monomer friction is \cite{LongPaper}

\begin{widetext}
\begin{eqnarray}
\label{EQ:zeta_mon}
\zeta_m & \approx & \frac{2}{3} (D\beta)^{-1} \rho g^2(d) \Biggl(\frac{1}{12}\pi N^2 d^2 R_g \Biggl[ 15\sqrt{2} + 40 \frac{d}{R_g} + 12\sqrt{2}\left(\frac{d}{R_g}\right)^2  \Biggr]  \nonumber \\
&& +\rho \pi N h_0 \frac{1}{3\sqrt{2}(R_g^2-2\xi_\rho^2)^2} \Biggl[ 12\sqrt{2}\xi_\rho^7 + 12 d^4 R_g^3 \left(1-2\left(\frac{\xi_\rho}{R_g} \right)^2\right) \nonumber \\
&& + 4\sqrt{2} d^3 R_g^4 \left( 5-14 \left(\frac{\xi_\rho}{R_g}\right)^2 + 2 \left(\frac{\xi_\rho}{R_g}\right)^4 \right) + 3 d^2 R_g^5 \Biggl( 5- 14 \left(\frac{\xi_\rho}{R_g}\right)^2 - 4\sqrt{2} \left(\frac{\xi_\rho}{R_g} \right)^5 \Biggr)  \nonumber \\
&& - 12\sqrt{2} e^{-\frac{2d}{\xi_\rho}} \xi_\rho^7 \left(1+\frac{d}{\xi_\rho} \right)^2 \Biggr] + 
 \rho^2 \pi h_0^2 \frac{1}{12(R_g^2-2\xi_\rho^2)^3} \Bigg[ 40 d^3 R_g^6 + 15\sqrt{2} d^2 R_g^7  \nonumber \\
&& - 24\sqrt{2}d^4 R_g^3 \xi_\rho^2 - 144 d^3 R_g^4 \xi_\rho^2 +  6 \sqrt{2} d^4 R_g^5 \left(2 - 9 \left(\frac{\xi_\rho}{d}\right)^2 \right) \nonumber \\
&& + 12 R_g^2 \xi_\rho^7 \left(4 \left(\frac{d}{\xi_\rho}\right)^3 - 7 \left(\frac{d}{\xi_\rho}\right)^2 + 9 \right) - 8\xi_\rho^9 \left(4 \left(\frac{d}{\xi_\rho}\right)^3 - 9 \left(\frac{d}{\xi_\rho} \right)^2 + 15 \right) \nonumber\\
&& - e^{-\frac{2d}{\xi_\rho}} 12 \xi_\rho^4 (d + \xi_\rho) \left( R_g^2 (d+3\xi_\rho)(2d+3\xi_\rho)-2\xi_\rho^2(2d^2+5\xi_\rho d + 5 \xi_\rho^2) \right) \Biggl] \Biggl) \  . 
\end{eqnarray}
\end{widetext}

\noindent This equation contains quantities that are well defined once the system of interest is selected. For example  $g(d)$ is the pair distribution function at contact, $\xi_{\rho}= R_g/(\sqrt{2}+2 \pi R_g^3 \rho/N)$ is the density fluctuation correlation length with $\rho$, $Rg$ and $N$ already defined in the text, and $h_0$ is defined as $h_0=h(k=0)=-(1-2\xi_\rho^2/R_g^2)/\rho_{ch}$.
The only physical quantity that needs to be determined is the hard-sphere diameter, $d$. This is defined once the Lennard-Jones potential is mapped onto an effective repulsive hard-sphere system as described later in this section of the paper.

In the soft colloidal representation, the friction coefficient is analytically calculated from the definition of the memory function as 
\begin{eqnarray}
\zeta_{\text{soft}} &\cong & (\beta/3) \rho_{ch} \int_0^{\infty}dt  
 \int d\mathbf{r} \int d\mathbf{r'} g(r) g(r') F(r) F(r')\, \mathbf{\hat r} \cdot \mathbf{\hat r'} \nonumber \\
&&\times \int d \mathbf{R} \, 
S(R;t) S(|\mathbf{r}-\mathbf{r'}+\mathbf{R}|;t) \, ,
\label{EQ:zeta_softMF}
\end{eqnarray}

\noindent where $\rho_{ch}=\rho/N$ is the chain density.  All the other physical quantities that appear in the integral, for example the pair distribution function, $g(r)$, and its related force, $F(r)$, are now defined in relation to the description of the polymeric liquid as a liquid of soft spheres, which are point particles interacting through a soft repulsive potential. Each of these quantities is analytical and calculated from the solution of the OZ equation with the hypernetted chain closure approximation.\cite{YAPRL} 

The regime of interest in our calculations is the diffusive limit where, in reciprocal space, the wave vector $k \ll 1/R_g$. Here the dynamic structure factor can be approximated as

\begin{equation}
S(k;t)\approx S(k)\ e^{-D t k^2}=(1+\rho_{ch} h_{soft}(k))\ e^{-D t k^2} \, ,
\label{EQ:SFSC}
\end{equation}

\noindent with the total correlation function of the soft-sphere representation is given in the limit of long chains ($N \ge 30$) by the approximated expression\cite{YAPRL}

\begin{equation}
h_{soft}(r )\approx-\frac{39}{16}\sqrt{\frac{3}{\pi}}\frac{\xi_\rho}{R_g}\left(1+\sqrt{2}\frac{\xi_\rho}{R_g}\right) 
\left(1 - \frac{9 r^2}{26 R_g^2} \right) e^{-\frac{3 r^2}{4 R_g^2}} \, .
\label{EQ:2}
\end{equation}

The resulting equation for soft particle friction is \cite{LongPaper}

\begin{eqnarray}
\zeta_{soft} \cong 4 \sqrt{\pi} (D \beta)^{-1} \rho_{ch} R_g \xi_\rho^2 \left(1+\frac{\sqrt{2}\xi_\rho}{R_g} \right)^2 \nonumber \\
 \times\frac{507}{512} \left[ \sqrt{\frac{3}{2}}  + \frac{1183}{507} \rho_{ch} h_0 + \frac{679\sqrt{3}}{1024} \rho_{ch}^2 h_0^2 \right] \, ,
\label{EQ:zeta_soft}
\end{eqnarray}

\noindent where the value of each physical parameter (temperature and density) and molecular parameters (degree of polymerization and radius-of-gyration) that enter this equation is defined once the system to simulate is selected. These parameters are identical to the ones that are input to the equation of the monomer friction, Eq.(\ref{EQ:zeta_mon}), given that the two equations are representations for different levels of coarse-graining of the same system.

When the two equations for the monomer and the soft colloidal friction coefficients are introduced in Eq.(\ref{EQ:rescaling}) the diffusion coefficient for a single chain in a liquid is recovered as 
\begin{equation}
\label{EQ:diff}
D=D^{MS}\zeta_{soft}/(N\zeta_m R_g \sqrt{3m N /(2 k_B T)}) \, ,
\end{equation}
where we have assumed that the diffusion coefficients that appear in Eq.\ref{EQ:zeta_mon} and Eq.\ref{EQ:zeta_soft} are identical, as the long-time relaxation of the dynamic structure factor in both CG descriptions is guided by molecular diffusion.

To summarize, as the first step the thermodynamic and molecular structure, i.e. the radius of gyration or equivalently the persistence length, of a polymer melt has to be defined. Then, MS MD simulations are performed for a liquid of point particles interacting through a soft pair potential, $\beta v_{soft}( r)$, as described in several of our previous papers, where
\begin{equation}
\beta v_{soft}(r)=h_{soft}(r)-\ln [1+h_{soft}(r)]-c_{soft}(r) \ ,
\label{EQ:v(r)}
\end{equation}
with
\begin{equation}
c_{soft}(k)=h_{soft}(k)[1+\rho_{ch}h_{soft}(k)]^{-1} \ .
\end{equation}
The mean-square-displacement of the soft spheres is measured from the MS MD trajectories as a function of time (dimensionless), and the resulting diffusion coefficient is rescaled following Eq.\ref{EQ:diff} to obtain the reconstructed diffusion coefficient.

For the solution of the rescaling factors in Eq.\ref{EQ:diff} the only parameter that has to be defined is the value of the effective hard sphere diameter, $d$, in Eq.\ref{EQ:zeta_mon}. In our formalism the elementary interaction between monomers belonging to different chains, which in the UA MD is a Lennard-Jones potential, is approximated by a hard-sphere interaction with an effective diameter, $d$, to allow for the analytical solution of the monomer memory function.
Under fixed thermodynamic conditions, the value of $d$ should depend only on the local monomer structure, and be independent of the degree of polymerization, because the monomer interaction potential is a local property of the chain. Therefore in our model, once the value of $d$ is chosen,
this value is kept fixed for all samples with different molecular weights. Note that for convenience we use the term monomer to identify the $CH_x$ group, where $x=1,\, 2$ or 3.

To evaluate the value of $d$ we analyze the dimensionless quantity $D \zeta_m/(k_B T)$, which is identical to $ N^{-1}$ if the chain obeys strictly Rouse dynamics in the long-time limit. 
To fulfill this requirement the chain has to have $N$ much smaller than the entanglement degree-of-polymerization $N_e$ and larger than the value of $N=30$ for which chains start to obey Gaussian statistics, i.e. the central limit theorem applies.

\begin{figure}
 \centering 
\includegraphics[scale=1.00]{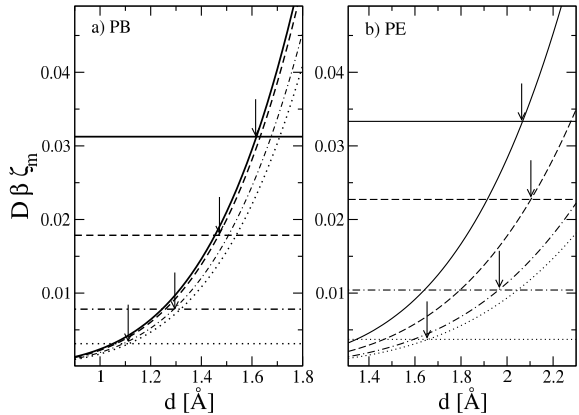}

 \caption[diffusion] {Dimensionless monomer friction coefficient as a function of the hard sphere diameter, based on Eq.(\ref{EQ:zeta_mon}). a) \textit{cis}-1,4-Polybutadiene samples with N= 32, 56, 128, 320 (solid, dashed, dot-dashed, and dotted lines correspondingly) and parameters as reported in Table  \ref{TB:UA_Rg} and  ref.~\cite{Tsolou}. b) Polyethylene samples with N=30, 44, 96, 270 (solid, dashed, dot-dashed, and dotted lines correspondingly) and parameters as reported in refs.~\cite{ShortPaper} and \cite{LongPaper}. Horizontal lines represent $1/N$ values, following the diffusion coefficient for unentangled chains, $D\beta\zeta_m = 1/N$.} 
\label{FG:zmon_of_d}
\end{figure}

Figure \ref{FG:zmon_of_d} a) displays the dimensionless friction coefficient, $D\beta\zeta_m$, from Eq.(\ref{EQ:zeta_mon}) as a function of the hard sphere diameter, $d$, for PB samples of increasing degree of polymerization, $N$. The parameter values ($\rho_m$, $T$, $N$, $R_g$) are chosen to match those of UA MD simulations in ref.~\cite{Tsolou} as reported in Table   \ref{TB:UA_Rg}. Horizontal lines represent the $1/N$ value for each sample, which is the Rouse value of the physical quantity $D\beta\zeta_m = 1/N$ in the diffusive regime, and reproduces the scaling behavior of the diffusion for unentangled, short-chains. For comparison Figure~\ref{FG:zmon_of_d} b) displays the analogous plot for the PE samples, which is reproduced from refs.\cite{ShortPaper} and \cite{LongPaper}.


The change of $d$ with $N$ observed in Figure \ref{FG:zmon_of_d} is a consequence of two different effects: for short chains, end effects are important, while for long chains the crossover to entangled dynamics starts to be important. In this way, it could be argued that in our method  the actual fitting parameter is not $d$, but the length of the chosen sample for which $d$ is determined. Clearly when entangled chains are forced to behave according to the Rouse expression the hard sphere diameter needs to be artificially decreased in order to compensate the overlooked increase in the monomer friction, which is associated with entanglement effects growing with increasing $N$. Whereas the range of lengths in the unentangled region can be rather wide, it is always possible to choose the chain length knowing experimentally measured or theoretically estimated value of $N_e$,\cite{entengl_Fetters} and by optimizing our choice of $d$ by testing the predictions of the theory against data from simulations of short chains. However comparison with simulations is not necessary, as the only needed information is the value of $N_e$ and the value of $R_g$ for a sample of short, unentangled, chains ($N < N_e$).\cite{LongPaper}

In our previous study of PE melts we fixed the hard sphere diameter following the same procedure described here, and we selected a sample with $N=44$ to calculate $d$. The obtained value of $d=2.1$\AA\ (see Figure \ref{FG:zmon_of_d}b), was larger than the carbon-carbon bond length $l_b=1.54$\AA, but smaller than the Lennard-Jones parameter $\sigma\simeq3.9$\AA\, typically used in UA MD simulations, which seems to be a reasonable choice of the parameter.
Compared to PE, PB chains are more flexible and so they are less entangled. The estimated entanglement degree-of-polymerization, $N_e$, is calculated from the generalized formula for the plateau value in the shear relaxation modulus,\cite{entengl_Fetters} which gives $N_{e(shear)}\simeq 330$ backbone carbon atoms for PB, compared to $N_{e(shear)} \simeq 55$ for PE. Notice that the value of $N_e$ from shear measurements is different from the value obtained from the analysis of scattering experiments which gives, for example, for PE samples $N_e \simeq 130$.

\begin{figure}
\centering 
\includegraphics[scale=1.00]{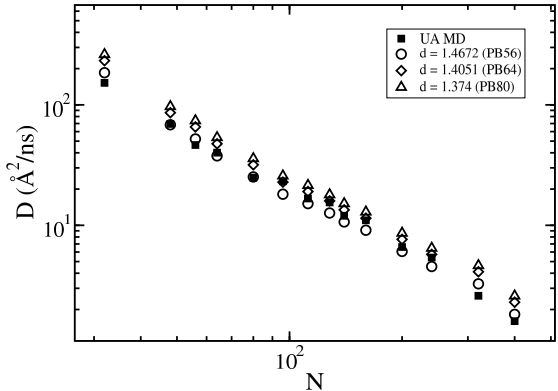}

\caption[diffusion] {Diffusion coefficient predicted from the rescaled MS MD, when different values of the hard-sphere diameter are chosen as an input to the reconstruction procedure. Assuming different values of $d$, calculated by enforcing Rouse diffusive behavior for N=56 (d=1.4672), 64 (d=1.4051), or 80 (d=1.374) (circles, diamond, triangles correspondingly) leads to diffusion coefficients in good agreement with the UA MD simulation data (filled squares).} 
\label{FG:dofn}
\end{figure}

It is important to notice that even if there is some freedom in choosing the reference sample, the deviations of $d$ from the chosen value is not large. However, small deviations of the value of $d$ can lead to different values of the diffusion coefficient. Figure \ref{FG:dofn} displays a comparison of the predicted diffusion coefficient as a function of $N$ when different values are chosen for the hard-sphere parameter $d$. It is possible to see that for values of $d$ obtained by enforcing Rouse dynamics for three different chain lengths that obey diffusive Rouse dynamics in the long-time regime, the predicted diffusion coefficients are all in reasonable agreement with the values of the UA MD simulation.

Notwithstanding the fact that for PB there is a wide range of degrees-of-polymerization, $30 < N < 330$, for which the hard-sphere diameter, $d$, could be optimized, the best agreement between theory and simulations is obtained for values of the parameter $d$ optimized for short chains, $N \approx 50$. This can be explained   by considering that in PB samples the crossover region from unentangled to entangled dynamics is extended (see for example Figure  \ref{FG:DofN_PB}, where deviations from the $N^{-1}$ scaling occur already for $N \approx 100$). Samples with $N$ approaching  $ N_e$ are not following the unentangled scaling behavior, which is necessary condition for the optimization of $d$ using Rouse-like diffusion. 
  
Considering the broader range of unentangeled PB chains, we select for the calculation of the hard sphere diameter the PB sample with $N=56$. The obtained value of $d\simeq1.47$\AA\ (which is smaller than the value for PE of $2.1$\AA\, given the flexibility of PB) is consistent with the fact that in PB chains half of the carbon bonds are shorter double bond and $l_{db} = 1.34$\AA\,.

\section{Theoretical predictions of center-of-mass diffusion}
The solution of Eq.(\ref{EQ:rescaling}) gives the diffusion coefficient for the center-of-mass of a polymer described at the atomistic level, having as an input the diffusion coefficient calculated from the mesoscale simulation of the CG polymer liquid. The purpose of this procedure is to have ultimately mesoscale simulations that, once properly rescaled, can provide directly the diffusion coefficient, without need of running atomistic simulations. To test our procedure we first run the mesoscale simulations, then we rescale the diffusion coefficient, and finally we compare the predicted values of $D$ against atomistic simulations and/or experiments. Thermodynamic and molecular parameters entering our equations have to be consistent with the parameters of the system against which we compare our approach.

We study the dynamics of \textit{cis}-1,4-Polybutadiene (PB) melts of increasing degree of polymerization and compare our results with the data from the simulations performed by Tsolou et al.\cite{Tsolou}  The input parameters to our theory are displayed in Table~\ref{TB:UA_Rg}, including the degree-of-polymerization, $N$, and the monomer density $\rho$. Table~\ref{TB:UA_Rg} also includes the mean-square-radius-of-gyration calculated from the united atom simulations $\left<R_g^2\right>^{UA}$. From this radius-of-gyration the value of the semiflexibility parameter, $g$, is derived, as the value that corresponds to chains with the desired overall  dimension, i.e. $R_g$. In this way the parameter $g$ is not an independent parameter but is determined by the $R_g$ value. The parameter $g$ does not enter the equation, but is used in the calculations reported in the last section of this paper. 
It is important to notice that the values for $R_g$ could be  taken from experimental data, and that, in principle, no atomistic simulations of the system under study are necessary for our approach.

The UA MD simulations were performed in the $NPT$ ensemble, at the constant  pressure, $P$, and temperature, $T$, reported in the table.\cite{Tsolou}
Note that the values of $R_g$ for systems PB160, PB200, PB240 were corrected from 257\AA$^2$ to 275\AA$^2$, from 304\AA$^2$ to 340\AA$^2$ and from 401\AA$^2$ to 410\AA$^2$ correspondingly, after consultation with the authors of ref.\cite{Tsolou}. The correction was motivated by the inconsistency observed with the data reported in reference \cite{Tsolou}  when the calculation of $R_g$ is performed using a standard model, such as the Freely-Rotating-Chain, described later in in this paper.

\begin{table}[h]
\centering
\caption{Simulation parameters for 1,4-cis PB chains of increasing lengths.}
\bigskip
\begin{tabular}{lccc}\hline \hline
\mbox{{\it System}}      
& $\rho$[sites/\AA$^{3}$] &  $\left<R_g^2\right>^{UA}$[\AA$^2$] &  g\\
\hline
PB  32	 &	 0.0352375	 &	 45$\pm$5	 	 &  0.6237\\
PB  48	 &	 0.0363767	 &	 70$\pm$5	 	 &  0.6243\\
PB  56	 &	 0.0367159	 &	 85$\pm$7	 	 &  0.6339\\
PB  64	 &	 0.0369744	 &	 95$\pm$10	 	 &  0.6246\\
PB  80	 &	 0.0373425	 &	 125$\pm$10	 	 &  0.6374\\
PB  96	 &	 0.0375921	 &	 152$\pm$16	 	 &  0.6394\\
PB  112	 &	 0.0377723	 &	 184$\pm$15	 	 &  0.6490\\
PB  128	 &	 0.0379087	 &	 215$\pm$18	 	 &  0.6544\\
PB  140	 &	 0.0379910	 &	 234$\pm$18	 	 &  0.6523\\
PB  160	 &	 0.0381012	 &	 275$\pm$25	 	 &  0.6595\\
PB  200	 &	 0.0382567	 &	 340$\pm$20	 	 &  0.6551\\
PB  240	 &	 0.0383610	 &	 410$\pm$30	 	 &  0.6557\\
PB  320	 &	 0.0384923	 &	 576$\pm$30	 	 &  0.6695\\
PB  400	 &	 0.0385714	 &	 678$\pm$30	 	 &  0.6519\\
\hline \hline
\multicolumn{4}{l}{T=413K, P=1 atm}\\
\end{tabular}
\label{TB:UA_Rg}
\end{table}

In the mesoscale simulations of a polymer melt, each chain is represented as a point particle interacting through a soft-core potential derived from the solution of the Ornstein Zernike equation applying the Hyper Netted Chain closure. MS MD simulations were implemented in the microcanonical $(NVE)$ ensemble on a cubic box with periodic boundary conditions. We used reduced units such that all the units of length were scaled by $R_g$ ($r^*=r/R_g$) and energies were scaled by $k_B T$. More details of our simulation procedure have been reported in previous papers.\cite{ShortPaper,jaypaper,multis}

Table \ref{TB:DofN_UA} reports the diffusion coefficient directly calculated from the MS MD in the soft sphere representation, $D^{\text{MS}}_{\text{t}}$,  and the diffusion coefficient calculated using the dynamical reconstruction procedure, $D$. Finally, for comparison, the Table shows the diffusion coefficient measured in the united atom simulation, $D^{UA}$.
The diffusion coefficient measured in the mesoscale simulations is several orders of magnitude faster than in the atomistic simulations. However, once it is rescaled the diffusion coefficient becomes very similar to the one obtained directly from the atomistic simulation. 
It is important to notice that, once the parameter $d$, which is characteristic of the polymer considered, is determined  the diffusion coefficient is calculated without any input from the dynamics of the atomistic simulations, so the procedure is predictive.

The samples here are relatively short and these calculations are performed for unentangled and slightly entangled systems. 
For strongly entangled systems, we proposed in a previous paper a perturbative version of our approach that accounts for the fact that both the tagged chain and the surrounding chains, that provide the entanglements, relax following the same dynamics.\cite{LongPaper}  In this manuscript the PB samples are in the unentangled and slightly entangled regimes, which are well represented by the theory without the one-loop perturbation.

\begin{table}[h]
\centering
\caption{Diffusion coefficient reconstructed from MS MD simulation compared against UA MD simulations.} 
\bigskip
\begin{tabular}{lccc}\hline \hline
\mbox{{\it N}}      
& $D_t^{MS}$[\AA$^2$/ns] & $D$[\AA$^2$/ns] & $D^{UA}$[\AA$^2$/ns]  \\
\hline
32	 &	 3875	 &	 184.5	 &	 152.2	 \\
48	 &	 3400	 &	 68.1	 &	 69.4	 \\
56	 &	 3425	 &	 51.9	 &	 46.4	 \\
64	 &	 3275	 &	 37.3	 &	 40.1	 \\
80	 &	 3174	 &	 25.1	 &	 24.9	 \\
96	 &	 3114	 &	 18.0	 &	 23.2	 \\
112	 &	 3235	 &	 15.1	 &	 16.7	 \\
128	 &	 3283	 &	 12.6	 &	 15.5	 \\
140	 &	 3221	 &	 10.5	 &	 12.0	 \\
160	 &	 2963	 &	 9.1	 &	 11.0	 \\
200	 &	 3245	 &	 6.0	 &	 6.6	 \\
240	 &	 3012	 &	 4.5	 &	 5.4	 \\
320	 &	 3386	 &	 3.2	 &	 2.6	 \\
400	 &	 3016	 &	 1.8	 &	 1.6	 \\
\hline \hline
\multicolumn{4}{l}{$d=1.4672$\AA (PB56)}\\
\end{tabular}
\label{TB:DofN_UA}
\end{table}

\begin{figure}
 \centering 
 \bigskip
\includegraphics[scale=1.0]{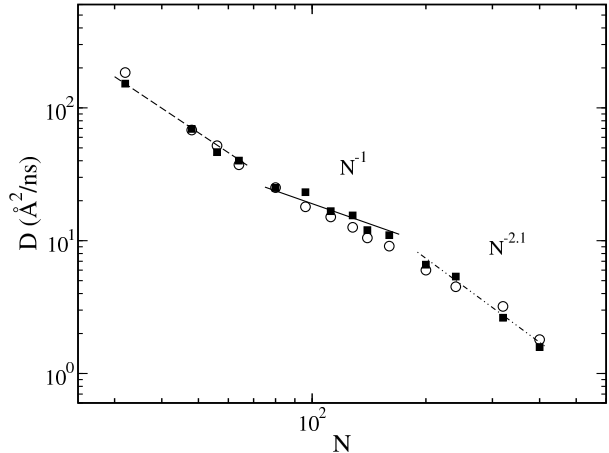}
 \caption[diffusion] {Center of mass self-diffusion coefficient as a function of degree of polymerization, N, for \textit{cis}-1,4-Polybutadiene melts with parameters defined in Table~\ref{TB:UA_Rg}. Diffusion coefficients reconstructed from MS MD by applying our procedure (open symbol) are compared against UA MD data (filled symbol) from reference \cite{Tsolou}. In analogy with the figure from the source, three scaling regimes in terms of power law dependence of $D \propto N^b$ are shown as dashed ($b> 1$), solid ($b \approx 1$), and dot-dot-dashed ($b \approx 2$) lines.}
\label{FG:DofN_PB}
\end{figure}

On Figure {\ref{FG:DofN_PB}} the diffusion coefficient is presented as a function of degree of polymerization $N$. Filled symbols represent UA MD simulations from ref. \cite{Tsolou} and open symbols represent MS MD simulations rescaled as in Eq.(\ref{EQ:rescaling}).
In analogy with the figure in Ref.\cite{Tsolou} from which the UA MD data are taken, we report three scaling regimes in terms of power law dependence of $D\propto N^{-b}$.For $N < 80$, there is a faster than Rouse regime with $D$ with $b>1$, which is attributed to the free-volume effects due to chain ends which is significant only for very short chains \cite{FreeVolume1, FreeVolume2, FreeVolume3}. In the intermediate regime for $80 < N < 160$ a Rouse-like behavior can be observed where the scaling exponent $b$ is close to 1. For longer chains with $N > 200$ the value of $b\approx 2.1$ is typical of the crossover to reptation dynamics. 

\begin{figure}
 \centering
 \bigskip 
\includegraphics[scale=1.00]{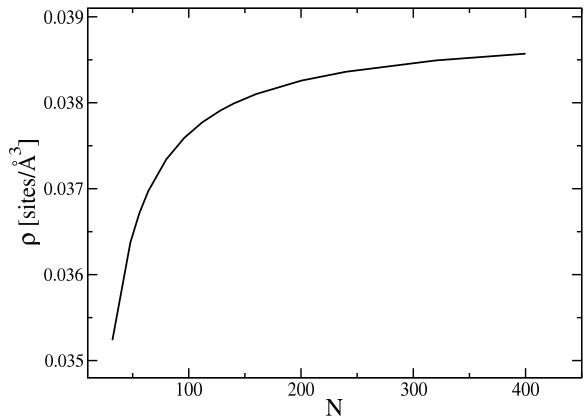}
 \caption[diffusion] {Density as a function of degree of polymerization, N, for \textit{cis}-1,4-Polybutadiene samples reported in Table~\ref{TB:UA_Rg}.}
\label{FG:dens_of_N}
\end{figure} 

The data in this plot are not closely following the reported scaling exponents because as the degree of polymerization increases the density of the system changes (Fig.\ref{FG:dens_of_N}). The different regimes are even less prominent in the MS MD simulations, where we observe a smoother transition between regimes than in UA MD data. In our calculations features like free volume effects due to chain ends (causing faster than Rouse decay of the diffusion coefficient for short chains) enter only indirectly as a consequence of the interplay of the input parameters, i.e. $\rho$, $R_g$, $N$. The increasing of the density with $N$ enters our analytical equation both directly, through the liquid density, and indirectly through the effective radius of gyration of the polymer, given that the compactness of the chain changes with the density (see also Section \ref{FRC} of this paper).

The \textit{cis}-1,4-PB chains are more flexible in comparison to PE chains, therefore we should expect larger entanglement length $N_e$ (for PE from scattering $N_e\simeq130$). Figure~\ref{FG:DofN_PB} shows that crossover from Rouse to reptation regime starts at $N=200$. The largest system shown on Figure~\ref{FG:DofN_PB} is for $N=400$ which is still a very weakly entangled system (1 or 2 entanglements per chain).
Overall the agreement between simulated and reconstructed diffusion coefficients is quite good.
  




\section{Internal dynamics}
\label{SX:Intnl}

The coarse-grained model adopted  in this paper simplifies the macromolecular structure by reducing the description of the molecule to an effective site that is centered on the center-of-mass of the molecule. For this reason, the only dynamics that can be predicted by this CG description, through the rescaling procedure, is the long-time diffusive behavior of the center-of-mass of the molecule. However from the knowledge of the diffusion coefficient the monomer friction coefficient can be obtained, and the latter can be used as an input to the Langevin equation that describes the internal dynamics of the polymer chain at any length scale of interest.

Specifically we follow our approach for the Cooperative Dynamics of a group of macromolecules, or Cooperative Dynamics Generalized Langevin Equation (CDGLE).\cite{Manychain} Conventional theories of polymer dynamics, such as the Rouse theory for unentangled polymer dynamics and the ``reptation" model for entangled dynamics, predict diffusive motion at short time following the ballistic regime; this is however in disagreement with simulations and experiments, which show a sub-diffusive regime before the com starts to follow Brownian motion. Both theories are mean-field approaches of single chain dynamics in an uniform bath.\cite{DoiEdw} Our approach  focuses on the dynamics of a group of polymer chains in a melt and  relates the anomalous subdiffusive behavior to the presence of cooperative motion of a group of polymer chains in a dynamically heterogeneous liquid,
as observed in simulations data of polymer melts and experiments.\cite{het1,het2,het3,het4} Theoretical predictions of the com and monomer  mean-square displacements are shown to be in quantitative agreement with computer simulations of unentangled\cite{MGuenza1,MGuenza2} and slightly entangled polymer melts,\cite{pallavi} and with scattering experiments of Neutron Spin Echo.\cite{Richter}

The physical picture underlying the theory builds on the fact that in polymer melts the dynamics appears heterogeneous with  the motion of a tagged chain correlated to the dynamics of a group of $n$ chains comprised inside the range of the intermolecular potential. The latter has a range of the order of the correlation hole, i.e. of the radius-of-gyration of the polymer, as it emerges from the solution of the OZ equation. The number of chains comprised in a volume of the order of $R_g^3$ is given by $n\approx \rho N^{1/2} l^{3}$, where $l$, the statistical segment length, and the other quantities have been defined earlier on. The number of interpenetrating chains, $n$, increases with increasing density, degree of polymerization, and the stiffness of the polymer. 

In the Cooperative Dynamic approach the dynamics is described by a set of coupled equations of motion (eom). Each equation is expressed in the space coordinates of the monomer $a$, belonging to molecule $i$ and positioned at $\mathbf{r}_a^{(i)}(t)$, and contains a balance of different forces acting on the monomer.
These include the viscous force, $\zeta \frac{d \mathbf{r}_a^{(i)}(t)}{dt} $, the
intramolecular force $- k_s \sum_{b=1}^N \mathbf{A}_{a,b} \mathbf{r}_b^{(i)}(t) $,which contains  the structural matrix $\mathbf{A}$, the time-dependent intermolecular mean-field force $\beta^{-1} \frac{\partial}{\partial\mathbf{r}_a^{(i)}(t)}
\ln \left[ \prod_{k < l}^n
g (\mathbf{r}^{(l)}(t),\mathbf{r}^{(k)}(t)) \right]$, and the random interactions with the
surrounding liquid, given by the random force
$\mathbf{F}_a^{(i)}(t)$.

\begin{eqnarray}
\label{seteom}
\zeta_{eff} \frac{d \mathbf{r}_a^{(i)}(t)}{dt} &  = & - k_s \sum_{b=1}^N \mathbf{A}_{a,b} \mathbf{r}_b^{(i)}(t)\nonumber\\
&&+\frac{1}{\beta}
\frac{\partial}{\partial\mathbf{r}_a^{(i)}(t)}
\ln \left[ \prod_{k < l}^n
g (\mathbf{r}^{(l)}(t),\mathbf{r}^{(k)}(t)) \right]\nonumber\\
&& + \mathbf{F}_a^{(i)}(t) \ .
\end{eqnarray}
Here ${\mathbf A}$ is the matrix of intramolecular connectivity, which reduces to the Rouse matrix when infinitely long and completely flexible macromolecules are considered, and is defined as 
\begin{equation}
\label{eqa}
A_{i,j}= \sum_{k,p=2}^N M_{k,i}U_{k,p}^{-1} M_{p,j} \ .
\end{equation}
Here ${\bf U}$ is the equilibrium averaged bond correlation matrix
\begin{eqnarray}
\label{equ}
U_{k,p}=\frac{<{\bf l}_k \cdot {\bf l}_p>}{l_k l_p}\ ,
\end{eqnarray}
and ${\bf M}$ is a structural matrix, with all the elements equal zero except $M_{1,i}=1/N$for $i=1, ..., N$, $M_{i,i}=1$ with $i=2, ..., N$, and 
$M_{i-1,i}=-1$ with $i=2, ..., N$. The ${\bf U}$ matrix is a function of the local semiflexibility parameter $g=< \mathbf{l}_i \cdot \mathbf{l}_{i+1}>/(l_il_{i+1})$, which is related to the persistence length of the polymer. For our samples the values of $g$ are reported in Table \ref{TB:UA_Rg} and are calculated from the molecule radius of gyration, which can be obtained from experiments or from atomistic simulations.\cite{LongPaper}

Through Eq.(\ref{equ}) the eom includes a complete microscopic description of the structure and local flexibility of the specific molecule under investigation, containing all the relevant parameters that define the intramolecular mean-force potential. 
Equations of motion for different beads belonging to the same chain or to two chains undergoing slow cooperative dynamics are coupled by the presence of intra- and inter-molecular interaction potentials. 
The intrinsic, chemical dependent semiflexibility of the macromolecule enters explicitly through the description of the ${\bf A}$ matrix.

The intermolecular potential is time dependent, as it  is a function of the relative position coordinates of the centers-of-mass of a pair of molecules. As the two molecules move with respect to each other the intermolecular force changes. From the initial ensemble of $n$ dynamically correlated chains, molecules diffuse in time and finally move outside the range of the potential, $R_g$, in a timescale of $\tau_{decorr}=R_g^2/D$.

The set of coupled equations is solved after transformation into normal modes of motion and numerically using a self-consistent procedure that calculate the effective potential at any given distance. Once $n$ and $\zeta_m = k_B T/(N D)$ are defined, the first optimized against the simulation data and the second from our rescaling procedure, the set of equations has no adjustable parameters. 

By applying consideration of symmetry the set of coupled eoms in normal coordinates can be simplified and reduced to two sets of $N$ uncoupled equations in the relative and collective normal-mode coordinates .
We assume that the eoms for internal modes ($p > 0$) do not contain intermolecular forces. This approximation is justified on the basis that polymer local dynamics is affected in a similar way by inter- and intra-molecular excluded volume interactions, which in polymer liquids tend to compensate each other. Intermolecular forces, which enter the dynamics through the eom for the first ($p=0$) normal mode, still perturb the dynamics on the local scale through the linear combination of the normal modes. 


In the following, we present an overview of some of the quantities analyzed in the original simulations\cite{Tsolou} and their comparison with the predictions of the Cooperative Dynamics theory having as an input parameter the reconstructed friction coefficient calculated from the soft-sphere simulations and the rescaling procedure as described in the first part of this paper.

\begin{figure}
 \centering 
\includegraphics[scale=1.0]{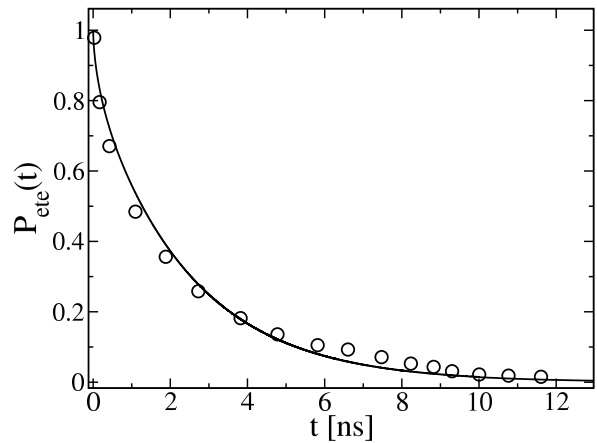}
 \caption[diffusion] {End-to-end vector time decorrelation function for \textit{cis}-1,4-Polybutadiene with N=112. Data from UA MD simulations (symbols) are compared with predictions of the theory for Cooperative Dynamics (solid line) where the number of correlated chains is set to $n=7$ and the monomer friction coefficient is reconstructed from MS MD simulations, using the procedure described in this paper.}
\label{FG:PetePB112}
\end{figure}

Figure~\ref{FG:PetePB112} presents the end-to-end vector, time decorrelation function for the $N=112$ sample of PB. The Cooperative Dynamic theory with the reconstructed monomer friction coefficient is directly compared against UA MD data. This function represents the rotational relaxation of the chain and cannot be measured from the soft colloid simulation directly. The $N=112$ sample has chains shorter than the entanglement length and the Cooperative Dynamics theory reproduces the atomistic simulation data rather well.

\begin{figure}
\centering 
\includegraphics[scale=1.0]{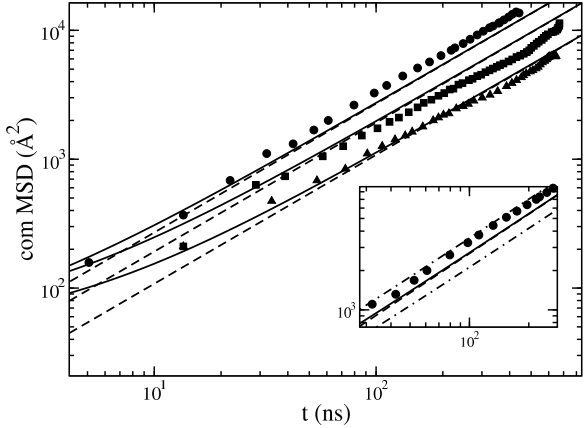}
\caption[diffusion] {Center of mass mean square displacement as a function of time for \textit{cis}-1,4-Polybutadiene samples with N=240 (circle), 320 (square), and 400 (triangle). Predictions of the theory for cooperative dynamics (solid lines) are compared against the UA MD simulations (symbols). Also shown are the purely diffusive slopes obtained  from the rescaled MS MD simulations (dashed lines). Inset illustrates how the uncertainty in the radius-of-gyration affects the mean-square-displacement for the N=240 sample: the upper and lower values, reported as dot-dashed lines, corresponds to the upper and lower values of $R_g$ as reported in Table \ref{TB:UA_Rg}.}
\label{FG:comMSD}
\end{figure}

Figure~\ref{FG:comMSD} shows the center of mass mean square displacement (com MSD) for three samples of \textit{cis}-1,4-Polybutadiene melts with chains of increasing degree of polymerization, $N=240, 320$ and $400$. The long time diffusive dynamics is calculated from the diffusion coefficients obtained from the rescaling procedure. For time shorter than the longest relaxation time, $\tau\approx R_g^2/D$, the mean-square displacement of the com shows a subdiffusive behavior that cannot be reproduced by the Rouse approach, i.e. the conventional theory of unentangled chain dynamics. In the cooperative dynamics theory the subdiffusive dynamics is a consequence of the cooperative motion of the interpenetrating chains inside the correlation hole region, coupled by the effective intermolecular potential between the coms of the macromolecules. The theory, with the  monomer friction from rescaled MS MD simulations as an input, is compared against the UA MD simulation data. The diffusion coefficient for PB240 is underestimated and for PB320 is slightly overestimated by the theory in comparison with the simulations, which is also can be seen in Table~\ref{TB:DofN_UA} and Figure~\ref{FG:DofN_PB}. These deviations are within the error related to $R_g$. Dot-dash lines in the inset of Figure~\ref{FG:comMSD} represent results calculated with the upper and the lower values of $R_g$ taken from the data of the atomistic simulations, as reported in Table~\ref{TB:UA_Rg} for the PB240 sample. The number of correlated chains is optimized to reproduce the subdiffusive behavior giving $n=10,10,12$ for PB240, PB320 and PB400 respectively, with a trend that qualitatively agree with the predicted scaling behavior. The error in the radius-of-gyration does not allow for a more precise calculation of the parameter $n$.

Starting from the theory of cooperative dynamics, the single chain dynamic structure factor can be calculated as
\begin{eqnarray}
S(\mathbf{q},t) &=&\frac{1}{N} \sum_{i,j}^N \left<exp [i \mathbf{q}\cdot \left(\mathbf{r}_i(t)-\mathbf{r}_j(0)  \right) ] \right> \nonumber\\
&\approx & \frac{1}{N} \sum_{i,j}^N exp \left[ -\frac{\mathbf{q}^2}{6} \left< \left(\mathbf{r}_i(t)-\mathbf{r}_j(0)  \right)^2 \right> \right] \ .
\label{EQ:skt}
\end{eqnarray}
Eq.\ref{EQ:skt} describes the dynamics of a "tagged" polymer chain, which moves in a group of $n$ interpenetrating chains. The intermolecular interaction between chains affects the dynamics resulting in the subdiffusive motion of the single-chain center-of-mass, as depicted in Figure \ref{FG:comMSD}, and in the subdiffusive decay of the low $q$ dynamics of $S(\mathbf{q},t)$. 

\begin{figure}
 \centering 
\includegraphics[scale=1.0]{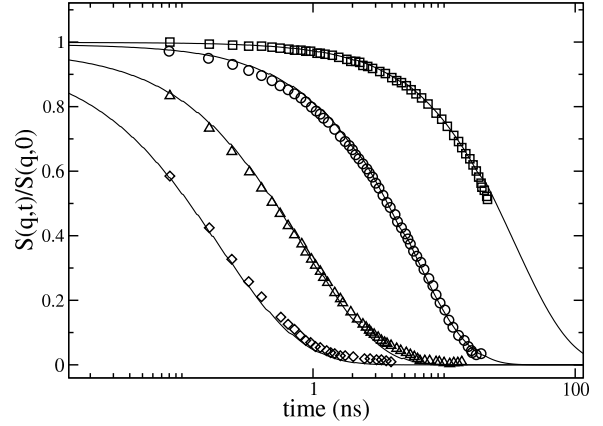}
 \caption[diffusion] {Normalized dynamic structure factor for \textit{cis}-1,4-Polybutadiene with N=96 and q=0.04 (squares), 0.1 (circles), 0.2 (triangles), 0.3 (diamonds) \AA$^{-1}$. The data from UA MD simulations (symbols) are compared against the Cooperative Dynamics theory (solid lines) where the number of correlated chains is set to $n=15$ and the monomer friction coefficient is reconstructed from MS MD simulations, using the procedure described in this paper. }
\label{FG:Sofk_PB96}
\end{figure}

\begin{figure}
 \centering 
  \bigskip
\includegraphics[scale=1.0]{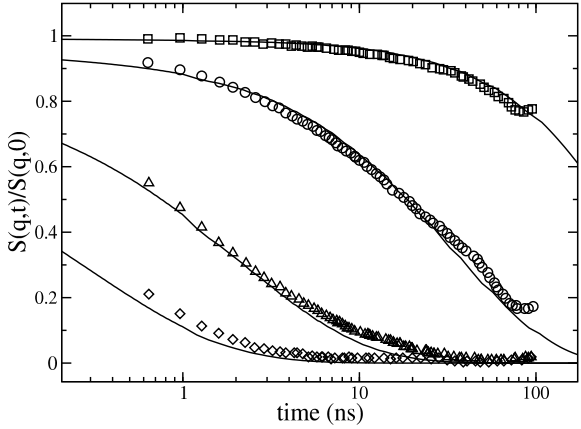}
 \caption[diffusion] {Normalized dynamic structure factor for \textit{cis}-1,4-Polybutadiene with N=400 and q=0.04 (squares), 0.1 (circles), 0.2 (triangles), 0.3 (diamonds) \AA$^{-1}$. The data from UA MD simulations (symbols) are compared against the theory for Cooperative Dynamics (solid lines) where the number of correlated chains is set to $n=12$ and the monomer friction coefficient is reconstructed from the MS MD simulations, using the procedure described in this paper. }
\label{FG:Sofk_PB400}
\end{figure}

The average time-dependent distance between two monomers in the tagged chain depends on the presence of the $n-1$ other chains dynamically correlated, and it is expressed as a function of relative and collective mode coordinates. 
Once $\zeta_{eff}$ and $n$ are determined, the theory predicts the full decay of the dynamic structure factor, at the different experimental wave vectors. Figures~\ref{FG:Sofk_PB96} and \ref{FG:Sofk_PB400} display the normalized dynamic structure factor obtained from the theory and  compared with UA MD simulations for $q=0.04, 0.1, 0.2, 0.3$ \AA$^{-1}$. For unentangled system PB96 shown on Figure~\ref{FG:Sofk_PB96} the agreement is excellent for all values of scattering vector $q$ and time range. The Rouse model (not shown in this figure, see ref. \cite{Tsolou}) fails to describe these data with satisfactory agreement, while the cooperative dynamics theory is found to be in quantitative agreement 
for samples at different molecular weights, over the whole time range, and for different wave vectors.

On Figure~\ref{FG:Sofk_PB400} the normalized dynamic structure factor is shown for the weakly entangled system PB400. The agreement for the smallest scattering vector which represents global dynamics is excellent at all times, while for intermediate and large values of $q$ the agreement between theory and simulations is slightly less accurate. The theoretical decay, which is slightly faster than simulations at long times, suffers from the lack of properly  describing the entanglement effects. A more recent version of the cooperative dynamics theory includes chain uncrossability, i.e. entanglements, and shows the correct slowing down of the relaxation in the long-time regime for entangled samples.\cite{marinainprogress} 

\begin{figure}
 \centering 
  \bigskip
\includegraphics[scale=1.0]{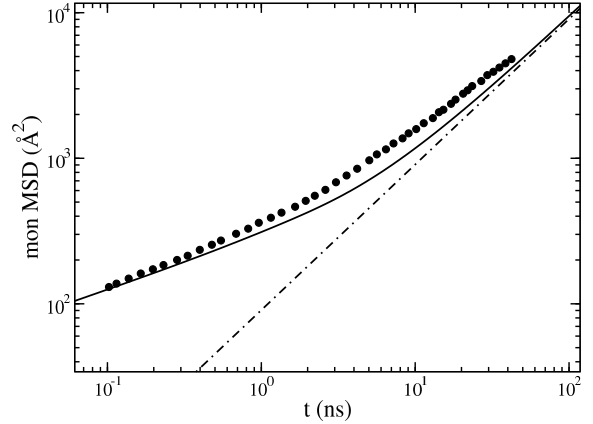}
 \caption[diffusion] {Monomer  mean square displacement, averaged over the innermost chain segments, as a function of time for \textit{cis}-1,4-Polybutadiene with N=112. Predictions from the theory of Cooperative Dynamics (solid line) are compared against UA MD simulations (circles). The slope obtained from the rescaled MS MD simulations is shown as well (dot-dashed line).}
\label{FG:monMSD}
\end{figure}

Figure~\ref{FG:monMSD} displays the monomer mean square displacement for \textit{cis}-1,4-Polybutadiene sample with $N=112$.
Theoretical prediction is compared against UA MD data from ref.~\cite{Tsolou}. The agreement in both subdiffusive and diffusive regions is very good. The theory accounts for the local semiflexibility and chain connectivity, together with the cooperative motion of the chains due to the presence of intermolecular interactions.

\section{Theoretical predictions using the freely rotating chain model}
\label{FRC}
The purpose of the rescaling procedure is to predict diffusion coefficients from the mesocale simulations properly rescaled when the atomistic simulations or the experiments are not available. One key input quantity to the procedure is the molecular radius of gyration. In the previous sections we adopted the values of $R_g$ determined from the atomistic simulations.
When there are no data available for the radius of gyration, neither from atomistic simulations nor from experiments, it is convenient to use some simple model, like the freely rotating chain model to provide $R_g$, as it captures the semiflexibility of the polymer chain.
In this section we present a critical discussion of adopting a freely rotating chain model to describe the PB chain, using a constant flexibility parameter $g$. 

For the \textit{cis}-1,4-polybutadiene chain with $N$ carbon atoms along its backbone there are $n_{CH=CH} = N/4$ double bonds with $l_{db} = 1.34$\AA, $n_{CH_2-CH_2} = N/4-1$ single bonds with $l_b=1.54$\AA \, and $n_{CH_2-CH} = N/2$ single bonds with $l_b' = 1.5$\AA. The average bond length over all $N-1$ bonds is calculated as

\begin{equation}
l_b^{ave} = \left[l_{db} \frac{N}{4} + l_b \left(\frac{N}{4}-1\right) + l_b' \frac{N}{2}\right]/(N-1)\, ,
\end{equation}

\noindent which in the $N\rightarrow\infty$ limit gives the value of $l_b^{ave} = 1.47$\AA.

The radius of gyration for a freely rotating chain (FRC) model with semiflexibility parameter $g$ is given as

\begin{equation}
R_{g}^2 = \frac{N-1}{6}\, (l_b^{ave})^2\, \left(\frac{1+g}{1-g}-\frac{2g}{(1-g)^2}\frac{1-g^{N-1}}{N-1}\right)  \, .
\label{EQ:FRC}
\end{equation}

\begin{figure}
 \centering 
\includegraphics[scale=1.0]{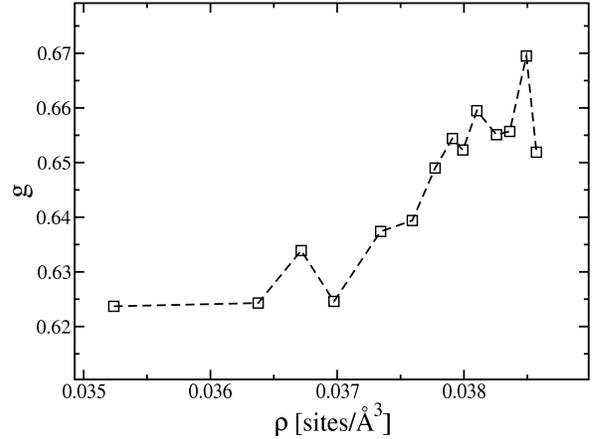}
 \caption[diffusion] {Density dependence of the semiflexibility parameter $g$ calculated from Eq.(\ref{EQ:FRC}) using values of the radius-of-gyration $R_g$ measured in UA MD and reported in Table~\ref{TB:g_paramPB}.}
\label{FG:g_param_of_rhom}
\end{figure}

\noindent Figure~\ref{FG:g_param_of_rhom} shows the density dependence of the semiflexibility parameter $g$ in the UA MD simulations.

\begin{table}[h]
\centering
\caption{Semiflexibility parameter $g$ calculated from FRC expression.}
\bigskip
\begin{tabular}{lcccc}\hline \hline
\mbox{{\it System}}      
&  $\left<R_{ete}^2\right>^{UA}$[\AA$^2$] & $g^a$  &  $\left<R_g^2\right>^{UA}$[\AA$^2$] &  $g^b$\\
\hline
PB  32  &   270$\pm$10   &   0.6237  &    45$\pm$5    &   0.6237  \\
PB  48  &   440$\pm$10   &   0.6391  &    70$\pm$5    &   0.6243  \\
PB  56  &   530$\pm$10   &   0.6458  &    85$\pm$7    &   0.6339  \\
PB  64  &   600$\pm$15   &   0.6406  &    95$\pm$10   &   0.6246  \\
PB  80  &   790$\pm$30   &   0.6531  &   125$\pm$10   &   0.6374  \\
PB  96  &   950$\pm$20   &   0.6516  &   152$\pm$16   &   0.6394  \\
PB 112  &  1150$\pm$20   &   0.6609  &   184$\pm$15   &   0.6490  \\
PB 128  &  1335$\pm$30   &   0.6643  &   215$\pm$18   &   0.6544  \\
PB 140  &  1430$\pm$40   &   0.6576  &   234$\pm$18   &   0.6523  \\
PB 160  &  1640$\pm$60   &   0.6578  &   275$\pm$25   &   0.6595  \\
PB 200  &  2100$\pm$80   &   0.6634  &   340$\pm$20   &   0.6551  \\
PB 240  &  2480$\pm$100  &   0.6581  &   410$\pm$30   &   0.6557  \\
PB 320  &  3100$\pm$200  &   0.6382  &   576$\pm$30   &   0.6695  \\
PB 400  &  4138$\pm$100  &   0.6568  &   678$\pm$30   &   0.6519  \\
\hline \hline
\multicolumn{5}{l}{$^a$ $R_{ete}=R_{ete}^{UA}$ ($g^{ave} = 0.651$); $^b$ $R_{ete}=\sqrt{6}R_g^{UA}$ ($g^{ave} = 0.645$)}\\
\end{tabular}
\label{TB:g_paramPB}
\end{table}

\begin{figure}
 \centering 
\includegraphics[scale=1.0]{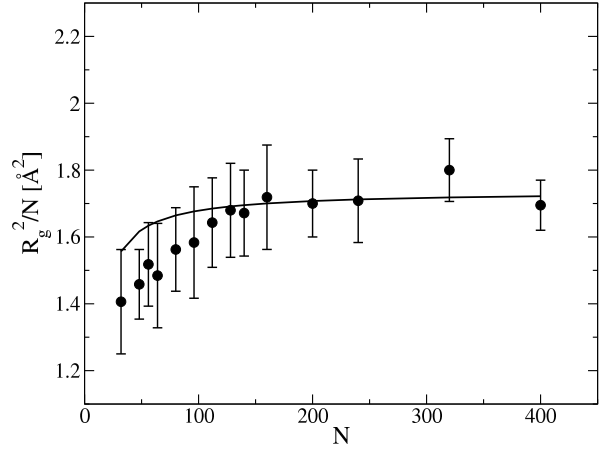}
 \caption[diffusion] {Radius of gyration squared over degree of polymerization $R_g^2/N$ as a function of $N$. UA MD data (circles), with statistical error, are compared against data calculated with the Freely Rotating Chain model using an averaged semiflexibility parameter, $g=0.6564$ (solid line).}
\label{FG:RgyrPB_of_N}
\end{figure}

Fitting simultaneously all samples to $R_g^{UA}$ data gives an average value of the semiflexibility parameter for all densities, $g=0.6564$. Figure~\ref{FG:RgyrPB_of_N} displays the radius of gyration squared over degree of polymerization $R_g^2/N$ as a function of $N$ from united atom simulations and from the freely-rotating-chain model. The curvature at small $N$ is due to chain-end effects. Samples with small $N$ present the largest deviation between the two sets of data, while the constant flexibility hypothesis works best for the large $N$ samples. The deviation is due to the fact that the united atom simulations are performed in the NPT ensemble where increasing $N$ corresponds to an increase of the liquid density, as shown in Figure~\ref{FG:dens_of_N}. The hypothesis of a constant semiflexibility parameter works well for the long chains for which coarse-graining methods are most useful. 

\begin{figure}
 \centering
\includegraphics[scale=1.00]{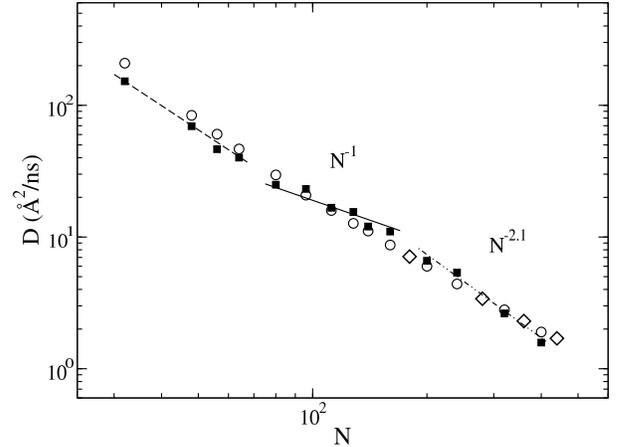} 
 \caption[diffusion] {Diffusion coefficients reconstructed from MS MD simulations using the radius-of-gyration calculated with the Freely Rotating Chain model (open circles) are compared against UA MD data (filled squares). Predictions for new systems with N=180, 280, 360, and 440 are shown as well (open diamonds).}
\label{FG:4}
\end{figure}

With this in mind, we report in Figure~\ref{FG:4} the diffusion coefficient calculated from the mesoscale simulations, properly rescaled, where now the input radius-of-gyration is taken from the freely rotating chain model. As it is expected the best agreement between UA simulations and theoretical predicted diffusion is for the samples with larger $N$. The disagreement at small $N$ is related to the fact that the density is changing with increasing $N$ in that region of the plot, and the radius of gyration should be properly corrected for this difference. In conclusion, if an exact value of the radius of gyration as a function of thermodynamic conditions and degree of polymerization is known, the described procedure should be able to provide an accurate estimate of the dynamics starting from the mesoscale simulations of the coarse-grained system, which are computationally very efficient. The figure also reports a number of samples for which neither experiments nor simulations have been performed ($N=180$, $280$, $360$, and $440$). The most relevant conclusion of this part of our study is the fact that it is possible to make predictions for data that are not reported in the UA MD simulations, obtaining information for new systems for which neither simulations nor experiments are available. 

\section{Conclusions} 
In a couple of recent papers we have presented an original approach to rescale the dynamics measured in mesoscale simulations of coarse-grained systems and recover the realistic dynamics as measured in atomistic simulations performed in identical thermodynamic conditions. The relevance of this method is in its ability of predicting diffusion coefficient, with good precision, directly from the simulation of a coarse-grained system, without the need of performing an atomistic simulation. Simply put, the method does not need a calibration curve. While the previous papers focused on the coarse-grained dynamics of polyethylene samples, this is the first paper in which the method is applied to a polymer with a different monomeric structure. 

In this paper we use an ``extreme" level of coarse-graining as the whole chain is represented as a soft colloidal particle. The choice of this representation is motivated by two reasons. The first is that a possible error in the procedure would be clearly made evident in this level of coarse-graining, and we believe that the procedure would produce an even better agreement if the macromolecule is coarse-grained at a lower level, e.g. as a collection of soft particles, or blobs.\cite{anthony,anthony1} The second is that for this representation all the physical quantities that enter our approach are analytically solved and the rescaling procedure is formally derived.

The paper presents a comparison between theoretical predictions of coarse-grained simulations properly rescaled following our first-principles procedure, and data from united-atom simulations for \textit{cis}-1,4-Polybutadiene melts. The simulations were performed by Tsolou et al. in a $NPT$ ensemble.\cite{Tsolou}
The rescaled diffusion coefficient shows good agreement with the diffusion measured in atomistic simulations, which supports the validity of the proposed procedure. 

As the coarse-graining model represents the whole polymeric chain as a soft colloidal particle, the only dynamical information that can be collected from the rescaled mesoscale simulation is the diffusion coefficient of the center-of-mass of the polymer. From this, however, the friction coefficient of the monomer is derived.
This paper shows how the monomer friction coefficient is used as an input to traditional Langevin equations for the dynamics of polymeric liquids. Specifically we use here the cooperative dynamics model for unentangled chains, given that the samples in the UA simulations are in the regime of unentangled and slightly entangled dynamics. From the theory of cooperative dynamics, which represents the cooperative motion of a group of interacting polymers in a dynamically heterogeneous liquid, all time correlation functions of interest can be calculated. Specifically we present here the comparison with the quantities reported in the original paper by Tsolou et al.\cite{Tsolou} and show that the theory, with the rescaled friction coefficient, is able to reproduce quantitatively the dynamics observed in the end-to-end time decorrelation function, and in the dynamic structure factor for unentangled and slightly entangled chains, and from the global to the local intramolecular dynamics. For entangled chains the local dynamics, represented by the high $q$ regime in the dynamic structure factor and in the monomer mean-square-displacement, shows a slight departure from the theory as the cooperative dynamics approach adopted here does not yet include the effect of entanglements.\cite{marinainprogress}

\section{Acknowledgements}
We acknowledge support from the National Science Foundation through grant DMR-0804145.
Computational resources were provided by Trestles through the XSEDE project supported by NSF. 
We are grateful to V. G. Mavrantzas for the informative discussion of his paper.



\begin{thebibliography}{99}
\bibitem{Binder} K. Binder Ed.   {\sl Monte Carlo and Molecular Dynamic Simulations in Polymer Science} (Oxford University Press, New York, 1995).
\bibitem{Frenkel} D. Frenkel, and B. Smit, \textit{Understanding Molecular Simulation} (Academic, New York, 2000).
\bibitem{shadowing} T. Sauer, C. Grebogi, and J. A.  Yorke Phys. Rev. Lett. \textbf{79}, 59 (1997). 
\bibitem{Tisley} M. P. Allen, and D. J. Tildesley  {\sl Computer Simulation of Liquids} (Oxford Science Publications, Oxford, 1992).
\bibitem{ShortPaper} I. Y. Lyubimov, J. McCarty, A. Clark, and M. G. Guenza, J. Chem. Phys. \textbf{132}, 224903 (2010).
\bibitem{LongPaper} I. Y. Lyubimov, M.G. Guenza Phys. Rev. E \textbf{84}, 031801 (2011).
\bibitem{Kroger} P. Ilg, H. C. \"Ottinger, and M. Kr\"oger Phys. Rev. E \textbf{79}, 011802 (2009).
\bibitem{Kroger1} P. Ilg,and M. Kr\"oger J. Rheol. \textbf{55}, 69 (2011).
\bibitem{Voth}Izvekov and Voth, J. Chem. Phys. 125, 151101 (2006).  
\bibitem{Padding} J. T. Padding, W. J. Briels J. Phys. Cond. Matt.  \textbf{23} 233201 (2011).
\bibitem{Klein} S. O. Nielsen, C. F. Lopez, G. Srinivas, M. L. Klein J. Phys. Condens. Mat. \textbf{16}, R481 (2004).
\bibitem{polymer} Y. Li, S. Tang, B. C. Abberton, M. Kröger, C. Burkhart, B. Jiang, G. J. Papakonstantopoulos, M. Poldneff, W. K. Liu Polymer \textbf{53}, 5935 (2012).
\bibitem{pb} J. T. Padding and W. J. Briels J. Chem. Phys. \textbf{117}, 925 (2002).
\bibitem{Vagelis} V. A. Harmandaris and K. Kremer Soft Matter \textbf{5}, 3920 (2009).
\bibitem{Tsolou} G. Tsolou, V. G. Mavrantzas, D. Theodorou Macromol. \textbf{38}, 1478 (2005).
\bibitem{Mavrantzas1} G. Tsolou, V.A. Harmandaris, V.G. Mavrantzas Macromol. Theory Simul. \textbf{15}, 381 (2006).
\bibitem{Mavrantzas2} P.S. Stephanou, C. Baig, G. Tsolou, V.G. Mavrantzas, and M. Kroger J.Chem.Phys \textbf{132}, 124904 (2010).
\bibitem{DoiEdw} Doi, M.; Edwards, S. F. {\sl The Theory of Polymer Dynamics} (Clarendon Press, Oxford, 1986).
\bibitem{Marina} M. G. Guenza J. Phys: Cond. Matt. \textbf{20}, 033101 (2008).
\bibitem{MGuenza1} Marina Guenza, Phys. Rev. Lett. \textbf{88}, 025901 (2002).
\bibitem{Richter} M. Zamponi, A. Wischnewski, M. Monkenbusch, L. Willner, D. Richter, P. Falus, B. Farago and M. G. Guenza J. Phys. Chem. \textbf{112}, 16220 (2008).
\bibitem{BJ} E. Caballero-Manrique, J. K. Brey, W. A. Deutschman, F. W. Dahlquist, and M. G. Guenza, Bioph. J. \textbf{93}, 4128 (2007).
\bibitem{entropy1} Y. Rosenfeld Phys. Rev. A \textbf{15}, 2545 (1977).
\bibitem{entropy2} J. A. Armstrong, C. Chakravarty, P. Ballone J. Chem. Phys. \textbf{136}, 124503 (2012).
\bibitem{Zwanzig}  R. Zwanzig \textit{Nonequilibrium Statistical Mechanics} (Oxford University Press, New York,  2001).
\bibitem{Schweizer} K. S. Schweizer, J. Chem. Phys. \textbf{91}, 5802 (1989).
\bibitem{YAPRL} G. Yatsenko, E. J. Sambriski, M. A. Nemirovskaya, and M. Guenza, Phys. Rev. Lett. \textbf{93}, 257803 (2004).
\bibitem{entengl_Fetters} L.J. Fetters, D.J. Lohse, C.A. Garcia-Franco, P. Brant, and D. Richter Macromol. \textbf{35}, 10096 (2002).
\bibitem{jaypaper} J. McCarty, I. Y. Lyubimov, M. G.  Guenza Macromol. \textbf{43}, 3964 (2010).
\bibitem{multis} J. McCarty, I. Y. Lyubimov, and M. G. Guenza J. Phys. Chem. B \textbf{113}, 11876 (2009).
\bibitem{FreeVolume1} F. Bueche, Physical Properties of Polymers (Interscience, New York, 1962).
\bibitem{FreeVolume2} V.A. Harmandaris, M. Doxastakis, V.G. Mavrantzas, D.N. Theodorou J.Chem. Phys. \textbf{116}, 436 (2002).
\bibitem{FreeVolume3} E. von Meerwall, S. Beckman, J. Jang, and W.L. Mattice J.Chem. Phys. \textbf{108}, 4299 (1998).
\bibitem{Manychain} M. G. Guenza, J. Chem. Phys. \textbf{110},7574 (1999).
\bibitem{het1} B. Frick, G. Dasseh, A. Cailliaux, and C. Alba-Simionesco  Chem. Phys. {\bf 292}, 311 (2003).
\bibitem{het2} A. Arbe, J. Colmenero, B. Farago, M. Monkenbusch, U. Buchenau, and D. Richter Chem. Phys. {\bf 292}, 295 (2003).
\bibitem{het3} J. Colmenero, F. Alvarez, and A. Arbe,  Phys. Rev. E {\bf 65}, 041804 (2002).
\bibitem{het4}  G. Tsolou, V. A. Harmandaris, and V. Mavrantzas  J. Non-Newtonian Fluid Mech. {\bf 152}, 184 (2008).

\bibitem{MGuenza2} Marina Guenza, Macromol. \textbf{35}, 2714 (2002).
\bibitem{pallavi} P. Debnath, and M. G. Guenza,   Phil. Mag. \textbf{88}, 33 (2008).
\bibitem{marinainprogress} M. G. Guenza (in preparation).
\bibitem{anthony} A. J. Clark, and M. G. Guenza, J. Chem. Phys. \textbf{132}, 044902  (2010).
\bibitem{anthony1} A. J. Clark,  J. McCarty, I. Y. Lyubimov, M. G.  Guenza Phys. Rev. Lett. \textbf{109}, 168301 (2012).
\end{thebibliography}
\end{document}